\documentclass[11pt]{iopart}
\pdfoutput=1
\expandafter\let\csname equation*\endcsname\relax
\expandafter\let\csname endequation*\endcsname\relax
\usepackage[numbers,sort&compress]{natbib} 
\usepackage[pdftex]{graphicx,color}
\usepackage{amsmath,amsfonts,amssymb,wasysym,graphicx,capt-of,ifthen,calc}
\usepackage{enumerate,float,url,color,subfigure}
\usepackage{amsmath}
\usepackage{graphics}  
\usepackage{psfrag} 
\usepackage[latin1]{inputenc}
\usepackage{color}
\usepackage{rotating}
\usepackage{url}
\usepackage{hyperref}
\begin{document}
\title{A Tale of Two (and More) Altruists}

\author{B. De Bruyne}
\address{LPTMS, CNRS, Univ.\ Paris-Sud, Universit\'e Paris-Saclay, 91405 Orsay, France}
\author{J. Randon-Furling}
\address{SAMM, Universit\'e Paris 1 -- FP2M (FR2036) CNRS, 75013 Paris, France}
\address{MSDA, Mohammed VI Polytechnic University, Ben Guerir 43150, Morocco}
\author{S. Redner}
\address{Santa Fe Institute, 1399 Hyde Park Rd., Santa Fe, New Mexico 87501, USA}


\begin{abstract}

  We introduce a minimalist dynamical model of wealth evolution and wealth
  sharing among $N$ agents as a platform to compare the relative merits of
  altruism and individualism.  In our model, the wealth of each agent
  independently evolves by diffusion.  For a population of altruists,
  whenever any agent reaches zero wealth (that is, the agent goes bankrupt),
  the remaining wealth of the other $N-1$ agents is equally shared among all.
  The population is collectively defined to be bankrupt when its total wealth
  falls below a specified small threshold value.  For individualists, each
  time an agent goes bankrupt (s)he is considered to be ``dead'' and no
  wealth redistribution occurs.  We determine the evolution of wealth in
  these two societies.  Altruism leads to more global median wealth at early
  times; eventually, however, the longest-lived individualists accumulate
  most of the wealth and are richer and more long lived than the altruists.

\end{abstract}

\section{Introduction}

Through the best of times and the worst of times, wealth inequality has
persisted in modern
societies~\cite{piketty2014inequality,piketty2015economics,gabaix2016dynamics}.
There is a long-standing debate about how to deal with this phenomenon.  An
individualistic viewpoint is that each person should try to maximize her or
his individual wealth without external constraints, and this opportunistic
perspective is best for society as a whole because those who thrive
economically can serve as wealth creators for others.  An extreme altruistic
viewpoint is that wealth should be shared equally by all.  Naively, altruism
would seem to forestall individual penury.  On the other hand, it might be
argued that wealth sharing stifles individual entrepreneurship, which
ultimately leads to less wealth for society as a whole.

We have no pretense that we are seriously addressing these vital issues;
instead, our goal is to investigate an idealized but solvable model of wealth
evolution and wealth sharing, for which we can make quantitative predictions
about the relative merits of altruistic and individualistic strategies.
Relevant investigations of this genre includes an unpublished note on the
so-called ``up the river'' problem by Aldous~\cite{aldous}, in which $N$
independent Brownian particles all start at $x = 1$ and are absorbed when
they hit $x = 0$.  A unit drift can be allocated arbitrarily among the
surviving particles to maximize the number of particles that survive forever.
The goal is to find the optimal allocation of the drift velocities for each
particle.  Related work by McKean and Shepp~\cite{mckean2006advantage}
investigate this same setting for two agents in which they determine the
optimal allocation of drift to maximize overall societal welfare, which is
related to the number of surviving agents.  A number of intriguing variations
of this basic problem have also been pursued~\cite{pal2008one,
  tang2018optimal,grandits2020ruin}.  The recent work of Abebe et
al.~\cite{abebe2020subsidy} is concerned with the optimal allocation of
subsidies to agents that experience negative economic shocks. A broader
perspective on the same type of question is that of optimal
tax-and-redistribution schemes~\cite{piketty2014optimal,chorro2016simple}.

There is also extensive literature on ruin problems in which the goal is to
understand the financial ruin of a single agent or firm that is subject to
positive and negative financial influences of various kinds (see, e.g.,
~\cite{cramer1955collective,andersen1957collective,thorin1982probabilities,dickson1998class}).
This perspective is often applied to the financial ruin of an insurance
agency: the positive financial influences are the premiums collected, the
negative ones are the claims paid to policyholders. Because this class of
problems involves a \emph{single} agent that responds to external influences,
there is a considerably better-developed analytical understanding of basic
phenomenology, and so-called ruin probabilities may be computed in models
beyond simple diffusion, such as L\'evy processes~\cite{asmussen2010ruin}.

While all these articles contain valuable insights, most of the problems
mentioned above typically involve considerable technical challenges.
Moreover, the nature of the optimal strategy depends on optimization
criterion itself.  In the work of McKean and Shepp, for example, the optimal
allocation of drift depends on whether the goal is to maximize the
probability that both agent remain solvent forever or maximize the expected
number of agents that remain solvent forever.

In this work, we construct an analytically tractable model of wealth
redistribution for which definitive conclusions can be reached with fairly
simple calculations.  In section \ref{sec:model} we introduce our altruism
model.  In the following two sections, we solve for the wealth dynamics of
two individualists and two altruists, respectively.  Then in section
\ref{sec:compare}, we determine whether altruism or individualism is superior
by studying both the time dependence of the survival probability---the
probability that an agent is still economically viable---and the typical, or
median, wealth of each agent.  We give some concluding comments in the final
section.

\section{Altruism Model}
\label{sec:model}

Our model of wealth evolution and redistribution is defined as follows:

\begin{enumerate}

\item Two diffusing particles are initially located at $x_0$ and at $y_0$ on
  the positive real axis.  Each particle represents a person and its
  coordinate represents the person's wealth.

\item The wealth of each person subsequently evolves by free diffusion.

\item If a person's wealth reaches 0, (s)he has gone bankrupt.  The other
  person---an altruist---immediately shares half of her/his current wealth
  with the bankrupt person.  The wealth of the two people again move by free
  diffusion until the next individual bankruptcy.  This cycle of events end
  when the total wealth shrinks to a specified small level that defines the
  bankruptcy of both agents.
  
\end{enumerate}
We may equivalently represent the wealth of the two agents as the diffusion
of a single effective particle in two dimensions that starts at $(x_0,y_0)$
and moves in the positive quadrant (Fig.~\ref{fig:traj}).  When both agents
are solvent, the effective particle satisfies the constraint that both
coordinates $x$ and $y$ are positive.  Whenever one coordinate hits 0, the
particle is reset to the main diagonal along a ray that is perpendicular to
this diagonal; this update rule corresponds to equal wealth sharing.  There
are several basic questions that we will address about this process:

\begin{figure}[ht]
  \centerline{ \includegraphics[width=0.425\textwidth]{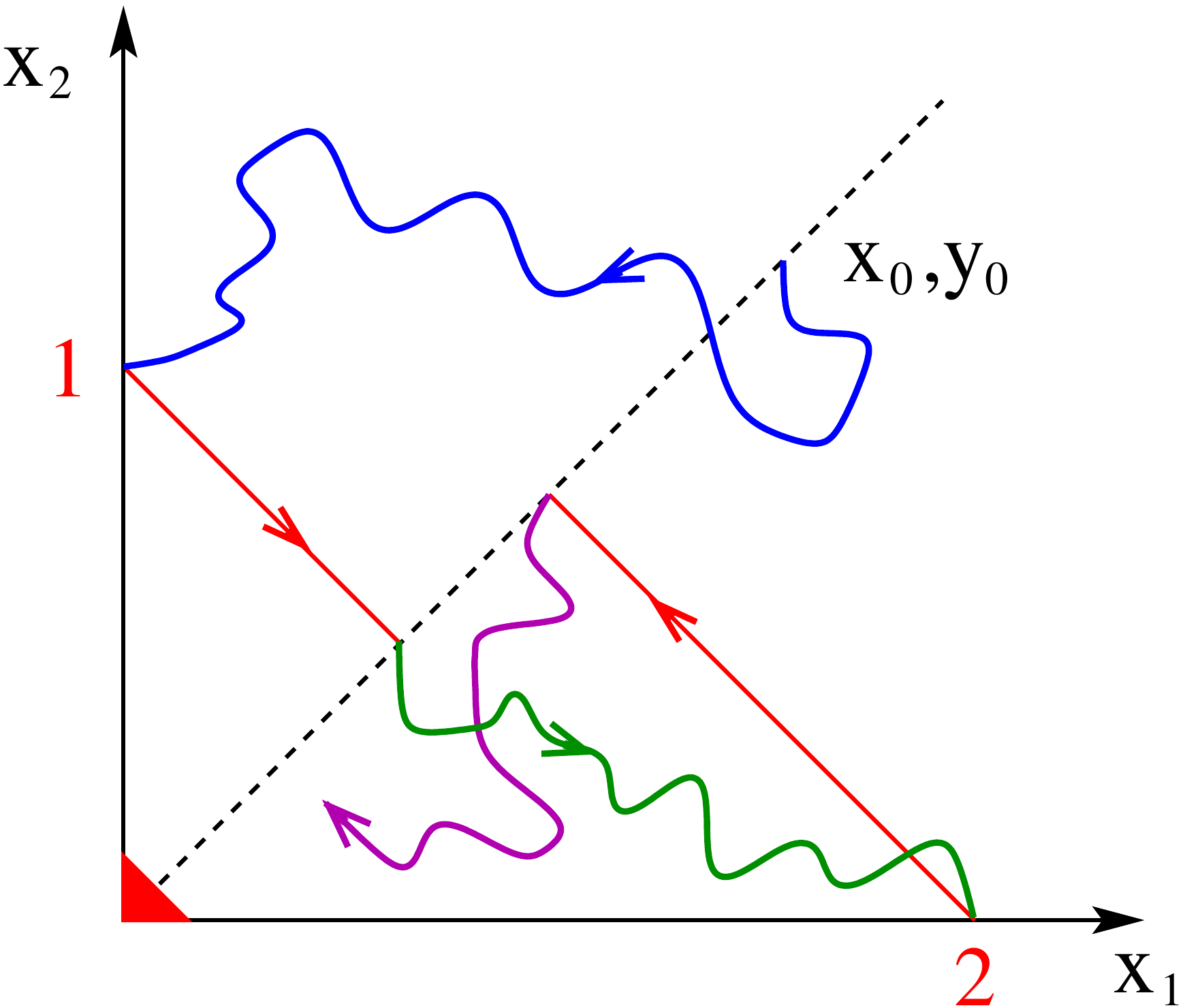}}
  \caption{Trajectory of the effective particle in two dimensions that
    represents the wealth of two altruistic agents.  Whenever a bankruptcy
    occurs (red numbers), where one of the coordinates reaches 0, the
    trajectory is reset to the main diagonal (dashed) along a ray
    perpendicular to the diagonal (red); this corresponds to equal-wealth
    sharing.  Joint bankruptcy occurs when the trajectory enters the small
    triangle near the origin. }
\label{fig:traj}
\end{figure}
\begin{enumerate}

\item When does the each individual bankruptcy occur?  At each such
  bankruptcy, how much wealth does the other individual have?
  
\item What is the wealth dynamics in successive bankruptcies?

\item What is the wealth dynamics for more than two people?

\end{enumerate}

In addition to answering these detailed questions about the dynamics, we are
also interested in comparing the evolution of wealth in this altruistic
society with that in an individualistic society.  In the latter, the wealth
of each individual also evolves by free diffusion, and whenever one person
does go bankrupt, (s)he is considered to be economically ``dead'', while the
wealth of the remaining $N-1$ solvent individuals continue to evolve by free
diffusion.  The existential question is: which society leads to a better
outcome?  As we shall see, the answer to this question depends on what is
defined as ``better''.  This same issue arises in other wealth sharing
models, where the notion of optimality depends on what is actually being
optimized~\cite{mckean2006advantage}.  We will first treat the simplest
non-trivial case of $N=2$ agents and then generalize to $N>2$.

\section{Dynamics of Two Diffusing Individualists}
\label{sec:ind}

As a preliminary, we solve for the wealth dynamics of two individualists.
Because diffusion in one dimension is
recurrent~\cite{feller2008introduction,redner2001guide,bray2013persistence}, one agent
necessarily goes bankrupt and subsequently the other necessarily goes
bankrupt. In spite of bankruptcy being certain for both agents, the average
time for each individual bankruptcy event is infinite.  To find the time of
the first bankruptcy, we again represent the wealth evolution as the motion
of an effective freely diffusing particle in two dimensions, with the
constraint that both $x$ and $y$ must be positive (Fig.~\ref{fig:traj}).  By
the image method~\cite{feller2008introduction,redner2001guide}, the
probability distribution of this effective particle is
\begin{align}
  P(x,y,t) &= \frac{1}{4\pi Dt}\,\left\{
  e^{-[(x-x_0)^2+(y-y_0)^2]/4Dt} - e^{-[(x+x_0)^2+(y-y_0)^2]/4Dt}\right.\nonumber\\
&\hskip 2cm  \left.- e^{-[(x-x_0)^2+(y+y_0)^2]/4Dt} + e^{-[(x+x_0)^2+(y+y_0)^2]/4Dt}\right\}\,.
\end{align}
This distribution is the sum of the initial Gaussian that starts at
$(x_0,y_0)$ plus the contribution of three image Gaussians---two negative
images at $(-x_0,y_0)$ and $(x_0,-y_0)$ and one positive image at
$(-x_0,-y_0)$---to enforce the condition that the probability vanishes on the
quadrant boundaries.

To find the time of the first bankruptcy, we compute the first-passage
probability to each boundary.  The first-passage probability to the
horizontal boundary, where the person with initial wealth $y_0$ goes bankrupt
first, is
\begin{subequations}
\begin{align}
  F_1(x,t|x_0,y_0) & = D\,\frac{\partial P(x,y,t)}{\partial y}\Big|_{y=0}\;.
\end{align}
That is, $F(x,t|x_0,y_0)$ is the probability that the first person goes
bankrupt at time $t$ and that the second person has wealth $x>0$ at the time
of this bankruptcy, when the initial wealth of the two individuals is
$(x_0,y_0)$.  For simplicity, we do not write this initial condition
dependence in what follows.  Performing the derivative and doing some simple
rearrangement gives
\begin{align}
  F_1(x,t)  & =\frac{y_0}{\sqrt{4\pi Dt^3}}\,e^{-y_0^2/4Dt}\times
                      \frac{1}{\sqrt{4\pi D t}} \left[ e^{-(x-x_0)^2/4Dt} - e^{-(x+x_0)^2/4Dt}\right]\,.\label{eq:F1}
\end{align}
\end{subequations}
This expression is just the product of the first-passage probability for the
$y$-coordinate to reach 0 times the probability that the $x$-coordinate
remains positive until this first passage.  Since these two events are
independent, the product of these probabilities gives the desired
first-passage probability.

Integrating over $x$, the distribution of times for the first of the two
agents to go bankrupt is
\begin{subequations}
\begin{align}
\label{phi2}
  \phi_1(t)  & =\int_0^\infty  F_1(x,t)\, dx
             = \frac{y_0}{\sqrt{4\pi Dt^3}}\,e^{-y_0^2/4Dt}\times \text{erf}(x_0/\sqrt{4Dt})\,.
\end{align}
Alternatively, this quantity is just the probability that the $y$ coordinate
first becomes zero at time $t$ multiplied by the probability that the $x$
coordinate always remains positive up to time $t$.  Using
$\text{erf}(z)\simeq 2z/\sqrt{\pi}$ for $z\to 0$, the long-time behavior of
$\phi_1(t)$ is
\begin{align}
  \label{phi-long}
  \phi_1(t)\simeq \frac{x_0y_0}{2 \pi Dt^2}\qquad t\to\infty\,.
\end{align}
\end{subequations}
As we might anticipate, the distribution of times for the first agent to go
bankrupt asymptotically decays faster than the same distribution for a single
agent ($t^{-2}$ versus $t^{-3/2}$~\cite{redner2001guide}).  Nevertheless, the
average time for this first bankruptcy event is still infinite.

By similar reasoning, the distribution of times for  the second person to
go bankrupt, $\phi_2(t)$, is
\begin{align}
  \label{F2}
  \phi_2(t) &= \int_0^\infty dx \int_0^t dt'\,
           \frac{y_0}{\sqrt{4\pi D  t'^3}}\; e^{-y_0^2/4Dt'} \times
           \frac{1}{\sqrt{4\pi
           Dt'}}\left[e^{-(x-x_0)^2/4Dt'}-e^{-(x+x_0)^24Dt'}\right]\nonumber\\
         &\hskip 6.3cm  \times  \frac{x}{\sqrt{4\pi D(t-t')^3}}\; e^{-x^2/4D(t-t')}\,.
\end{align}
The leading factor is the probability for the first person, with initial
wealth $y_0$, to go bankrupt at a time $t'<t$.  The second factor is the
wealth distribution of the second person at the moment of the first
bankruptcy.  This distribution defines the effective initial condition of the
second person.  The last factor is the probability that this second person,
with wealth $x$, goes bankrupt in the remaining time $t-t'$.  Performing the
integrals in \eqref{F2} gives
\begin{align}
  \phi_2(t) = \frac{x_0}{\sqrt{4\pi Dt^3}}\; e^{-x_0^2/4Dt}\; \text{erfc}(y_0/\sqrt{4Dt})\,.
  \label{eq:F2}
\end{align}
The complementary error function gives the probability that one person has
already gone bankrupt by time $t$, while the remaining factor gives the
probability of the other agent goes bankrupt at time $t$.  This joint
probability $\phi_2(t)$ has the same $t^{-3/2}$ asymptotic behavior as the
classic first-passage probability in one
dimension~\cite{feller2008introduction,redner2001guide,bray2013persistence}.  We also note that
$\phi_1$ and $\phi_2$ may be derived using order statistics: if one defines
$T_1$ and $T_2$ as the two independent random variables corresponding to the
first passage times to the origin of two independent Brownian motions, then
$\phi_1$ is the probability distribution of the minimum of $T_1$ and $T_2$
and $\phi_2$ is the distribution of the maximum.

\section{Dynamics of Two Diffusing Altruists}
\label{sec:alt}

We now incorporate altruism, in which after each individual bankruptcy, the
solvent person shares half of her/his wealth with the bankrupt person.  This
same wealth sharing rule occurs at each subsequent individual bankruptcy.  We
want to understand how such a wealth-sharing rule influences the time
dependence of the collective wealth.


To determine the wealth distribution of the solvent agent, $\mathcal{F}(x)$,
at the moment of the first bankruptcy, we integrate $F_1(x,t)$, the
time-dependent wealth distribution of the solvent agent, in Eq.~\eqref{eq:F1}
over time:
\begin{subequations}
\begin{align}
\label{Jx}
  \mathcal{F}(x) &= \int_0^\infty F_1(x,t)\, dt
  = \int_0^\infty  \frac{y_0\, e^{-y_0^2/4Dt}}{4\pi Dt^2}
  \left[e^{-(x-x_0)^2/4Dt} - e^{-(x+x_0)^2/4Dt} \right]\,dt\nonumber\\[3mm]
& = \frac{y_0}{\pi}\left[\frac{1}{y_0^2+(x+x_0)^2}-\frac{1}{y_0^2+(x-x_0)^2}\right]\,,
\end{align}
where the integral is performed by making the substitution $z=1/4Dt$.  The
probability $P_B$ that the person with initial wealth $y_0$ goes bankrupt first
is
\begin{align*}
  P_B &= \int_0^\infty \mathcal{F}(x)\, dx = \int_0^\infty dx\,
      \left[\frac{1}{y_0^2+(x+x_0)^2}-\frac{1}{y_0^2+(x-x_0)^2}\right]\nonumber\\[2mm]
  &  = \frac{2y_0}{\pi} \int_0^{x_0} \frac{dw}{w^2+y_0^2} = \frac{2}{\pi}\,
    \tan^{-1}(x_0/y_0)\,,
\end{align*}
where we make the substitution $w= x+x_0$ in the first integral and $w=x-x_0$
in the second.  When both people possess the same initial wealth, $x_0=y_0$,
the above formula gives the obvious result $P_B=\frac{1}{2}$, while for
$y_0=0$, $P_B=1$.

Henceforth we focus on the symmetric initial condition, $x_0=y_0$.  From
Eq.~\eqref{Jx} and multiplying by 2 to account for either person going
bankrupt first, the wealth distribution of the solvent person at the moment
of the first bankruptcy is
\begin{align}
\label{Jx-symm}
  \mathcal{F}(x) = \frac{1}{\pi}\, \frac{8 x x_0^2}{4x_0^4+x^4}\,.
\end{align}
\end{subequations}
Notice that $\int_0^\infty \mathcal{F}(x)\, dx$, which is the probability
that either person eventually goes bankrupt, equals 1, as it must.  The
average wealth of the solvent person at the first bankruptcy is
\begin{align*}
  \langle x\rangle &= \frac{\displaystyle \int_0^\infty x\, \mathcal{F}(x)\, dx}
                     {\displaystyle \int_0^\infty \mathcal{F}(x)\, dx}
  = \int_0^\infty \frac{8x^2x_0^2}{4x_0^4+x^4}\, dx = 2x_0\,.
\end{align*}
If this person now shares half of her/his wealth with the bankrupt person,
then both people restart with average wealth $x_0$.  This conservation of the
average wealth is a consequence of diffusion being a
martingale~\cite{morters2010brownian}.

However, this average outcome is not representative of a typical realization
of the dynamics.  When an individual bankruptcy occurs, the probability that
the wealth of the solvent person is less than $2x_0$ is
\begin{align*}
\int_0^{2x_0}  \mathcal{F}(x)\, dx=  \frac{2}{\pi}\tan^{-1}2\approx 0.7048\,.
\end{align*}
After sharing half of her/his wealth with the bankrupt person, the typical,
or median, wealth of each person will therefore be less than $x_0$.
Consequently, the typical wealth of the two people systematically decreases
to zero upon repeated bankruptcies.  This dichotomy between the typical and
average outcome arises because the wealth distribution of the solvent person
at the moment of bankruptcy asymptotically has an algebraic $x^{-3}$ tail
(Eq.~\eqref{Jx-symm}).  The average incorporates rare events where $x$ is
anomalously large, and these are absent in typical events.

To determine how the typical wealth decays with time, we rewrite
$\mathcal{F}(x)$ in \eqref{Jx-symm} as
\begin{align*}
  \mathcal{F}(x)\, dx =\frac{8}{\pi}\; \frac{x/x_0}{4+(x/x_0)^4}\; \frac{dx}{x_0}~.
\end{align*}
That is, the typical wealth of the solvent person is rescaled by
$x\to \eta x$ at each bankruptcy, where the probability distribution for the
random variable $\eta$ is
\begin{align*}
P(\eta)\,d\eta = \frac{8}{\pi} \frac{\eta}{4+\eta^4}\,d\eta\,.
\end{align*}
After the joint wealth is shared, the typical wealth of both people now is
$x=\frac{1}{2}\eta\, x_0$.  After the $n^{\text th}$ bankruptcy, each person
therefore has typical wealth $x_n= \frac{1}{2}\eta \,x_{n-1}$; namely, $x_n$
follows a geometric random walk.  In terms of $S_n=\ln x_n$ and
$\xi= \ln(\eta/2)$, $S_n$ follows a simple random walk in which $S_n=S_{n-1}+\xi$.
The distribution of the random variable $\xi$ is determined by the
transformation $f(\xi)\,d\xi = P(\eta)\,d\eta$, which leads to
\begin{align}
\label{fxi}
  f(\xi) =\frac{8}{\pi} \frac{e^{2\xi}}{1+4e^{4\xi}}\,,
\end{align}
which is properly normalized on $[-\infty,\infty]$.  The salient point is
that the average value of $\xi$ is given by
$\langle\xi\rangle =\int_{-\infty}^{\infty} \xi\,f(\xi)\,d\xi =
-\ln\sqrt{2}$, which means that the typical wealth of both people is reduced
by a factor $1/\sqrt{2}$ after each bankruptcy.

\begin{figure}[ht]
\subfigure[]{\includegraphics[width=0.45\textwidth]{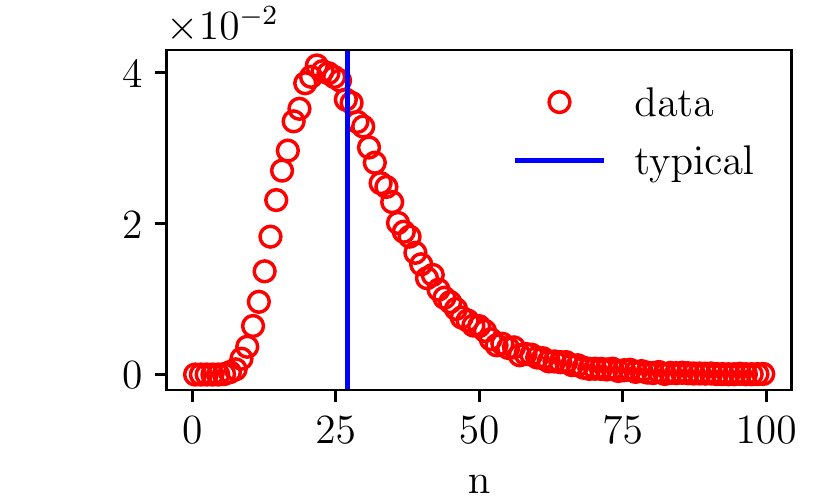}}\qquad
\subfigure[]{\includegraphics[width=0.45\textwidth]{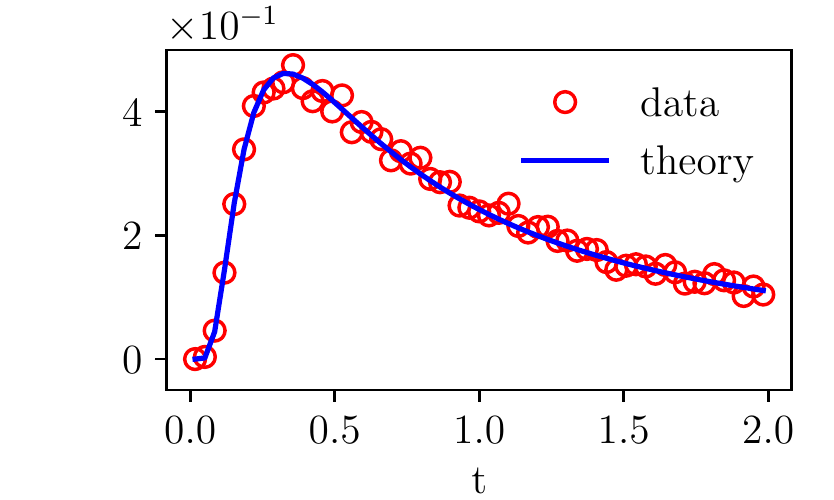}}
\caption{The probability distribution that joint bankruptcy for two altruists
  occurs after $n$ individual bankruptcy events (a) or after a time $t$ (b).
  The altruists start with initial wealth $x_0=1$.  In (a), the peak is at
  $n=22$, the median is at $n=27$, and $\langle n\rangle =29.14$. In (b), the
  peak is at $t=0.33$ and the median is at $t=2.21$.}
\label{fig:Qn}
\end{figure}

Because the typical wealth decreases multiplicatively after each bankruptcy,
the wealth of each person ultimately becomes vanishingly small.  Since this
wealth is always non-zero, we need to define the notion of joint bankruptcy
through a cutoff.  We postulate that joint bankruptcy occurs when the total
wealth of both people has been reduced by a factor of $10^{-4}$ (see
Fig.~\ref{fig:traj}).  At this level, we regard the wealth to be too small
for an agent to be economically viable; our main results are independent of
this cutoff, as long as it is sufficiently small.  The number of individual
bankruptcies needed to reach this state of joint bankruptcy is determined by
$(1/\sqrt{2})^n = 10^{-4}$ or $n\approx 27$. This prediction agrees with
simulations of the time evolution of the joint wealth in
Fig.~\ref{fig:Qn}(a).

We now determine how long it takes for both altruists to be bankrupt when
they both start with the same wealth.  It is useful to focus on the
longitudinal coordinate along the main diagonal,
$X_a(t) = \frac{1}{2}[x_1(t)+x_2(t)]$, which is just the average wealth of
the two altruists.  The diffusion coefficient associated with this
coordinate, $D_\parallel$, is related to $D$ by $D_\parallel=\frac{1}{2}D$.
The factor of $\frac{1}{2}$ occurs because the component of a wealth
increment or decrement of either agent is reduced by a factor $1/\sqrt{2}$
when this displacement is projected onto the main diagonal.

By construction, wealth redistribution after a bankruptcy does not affect
$X_a(t)$.  Thus $X_a(t)$ is simply a one-dimensional Brownian motion that
starts at $x_0$ in the presence of an absorbing boundary at $10^{-4}\,x_0$;
we approximate the location of this boundary by 0.  Thus the first-passage
probability for the altruists to reach the cutoff is
\begin{align}
  \label{eq:Falt}
  F_a(x_0,t) = \frac{1}{\sqrt{4\pi D_\parallel\,t}}\;e^{-x_0^2/4D_\parallel t}\,.
\end{align}
The numerical data shown in Fig.~\ref{fig:Qn}(b) for the first-passage
probability to joint bankruptcy accurately fits \eqref{eq:Falt}.  Note that
if the altruists employed an \emph{unequal} sharing rule, the first-passage
probability \eqref{eq:Falt} will remain valid because the coordinate $X_a(t)$
is unaffected by an unequal redistribution rule.

The results for two altruists can be straightforwardly generalized to more
altruists, and we quote some basic results for three altruists with the
symmetric initial condition $x_0=y_0=z_0=1$.  The analogue of \eqref{Jx-symm}
for the wealth distribution of the two solvent agents when the third agent
goes bankrupt is
\begin{align}
\label{Jx-symm-3d}
  \mathcal{F}(x) =  \frac{1}{\pi}\left\{\frac{1}{[(x\!-\!2) x+2] \sqrt{(x\!-\!2)
  x+3}}-\frac{1}{[x (x\!+\!2)+2] \sqrt{x (x\!+\!2)+3}}\right\}\,.
\end{align}
This distribution has the same $x^{-3}$ tail as in the case of two altruists.
From this distribution, the average wealth of the two solvent agents when the
other agent goes bankrupt is 3/2.  After wealth sharing, each agent again has
average wealth equal to 1.  However, the typical wealth decreases after each
bankruptcy by a factor of approximately 0.8022.  With this reduction factor,
the number of individual bankruptcies for the total wealth to be reduced by
$10^{-4}$ is now 42.  As the number of agents increases, the number of
individual bankruptcies before collective bankruptcy occurs increases
commensurately.

\section{Which is Better:  Altruism or Individualism?}
\label{sec:compare}

We now address the fundamental question of whether altruism or individualism
is better.  Specifically, are altruists more likely to ``live'' longer than
individualists?  Here the term ``live'' means that each altruist possesses
sufficient wealth (greater than the cutoff) to be economically viable.  A
related question is: who has more wealth---the altruists or the
individualists?  We first address these questions for two agents, and then
extend our considerations to any number of agents.

\subsection{Two Agents}

A basic ingredient in the following is $S(x_0,t,D)$, the survival probability
of a one-dimensional Brownian motion in the presence of an absorbing boundary
at the origin~\cite{redner2001guide}:
\begin{align}
  S(x_0,t,D) = \text{erf}\left(\frac{x_0}{\sqrt{4\,D\,t}}\right)\,.\label{eq:S}
\end{align}
Since the wealth of the two altruists is also described by one-dimensional
Brownian motion with diffusion coefficient $D_\parallel=D/2$, the ultimate
survival probability of the altruists is
\begin{align}
\label{eq:Salt}
  S_a(t) = S(x_0,t,D_\parallel)\,.
\end{align}

\begin{figure}[ht]
\centering \includegraphics[width=0.6\textwidth]{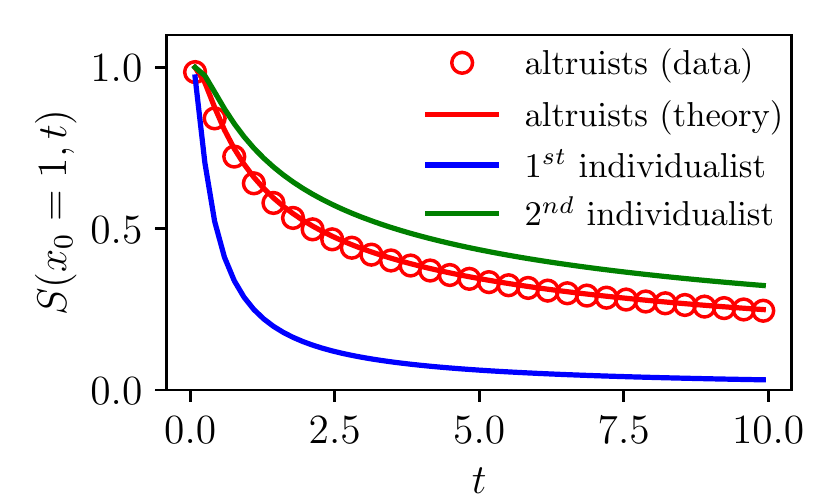}
\caption{The survival probabilities in a society of $N=2$ agents.  The
  altruists (red), and the first and second individualists to go bankrupt
  (blue and green, respectively).  Both agents start with wealth $x_0=1$. }
\label{fig:S}
\end{figure}

The survival probability of the first individualist, which is the probability
that both individualists are still alive, is
\begin{subequations}
\begin{align}
  \label{eq:S1}
  S_1(t) = \big[S(x_0,t,D)\big]^2\,.
\end{align}
The survival probability of the second individualist equals the probability
that both individuals are alive plus the probability that one is alive:
\begin{align}
  \label{eq:S2}
  S_2(t) = S(x_0,t,D)^2 + 2\,[1-S(x_0,t,D)]S(x_0,t,D)\,.
\end{align}
\end{subequations}
As shown in Fig.~\ref{fig:S}, the individualist that goes bankrupt first has
the worst possible outcome, while the individualist that goes bankrupt second
has the best outcome.  The survival probability of the altruists is
intermediate to those of each individualist.

To determine the time dependence of the altruists' and individualists'
wealth, we need $G(x,t)$, the propagator of the fraction of surviving
one-dimensional Brownian motions that start at $x_0$ in the presence of an
absorbing boundary at the origin:
\begin{align} 
  \label{eq:G}
  G(x,t,D) =   \frac{1}{\sqrt{4\pi D t}}\left(e^{-(x-x_0)^2/4Dt}
  -e^{-(x+x_0)^2/4Dt}\right)\bigg/\text{erf}\left(\frac{x_0}{\sqrt{4Dt}}\right)\,.
\end{align}
The error function normalizes this propagator so that its spatial integral
equals 1.  

\begin{figure}[ht]
  \centering \includegraphics[width=0.7\textwidth]{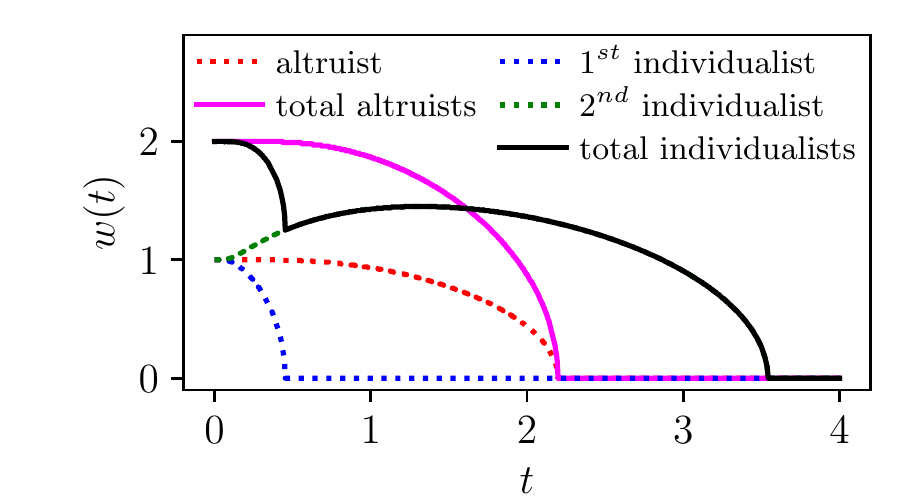}
  \caption{Evolution of the typical wealth in a society of $N=2$ agents.  The
    altruists (red), the first individualist to go bankrupt (blue), the
    second individualist (green), the altruistic society (magenta), and the
    individualistic society (black).  Both agents have initial wealth
    $x_0=1$.}
\label{fig:xt}
\end{figure}

For two altruists, their wealth distribution is
\begin{align}
  G_a(x,t) = S_a(t)\,G(x,t,D_\parallel)\,.
\end{align}
A natural way to quantify the wealth of the agents is by its \emph{typical}
or \emph{median} value.  The median wealth for the altruists, $w_a(t)$, is
obtained from the criterion
\begin{align}
  \label{eq:typg1}
  \int_{w_a(t)}^\infty\,dx\,G_a(x,t) = \tfrac{1}{2}\,.
\end{align}
Namely, at the median wealth, one-half of the wealth distribution exceeds
$w_a$ and one-half is less than $w_a$.  This criterion leads, after a
straightforward integration over $x$, to the following implicit equation for
$w_a(t)$:
\begin{align}
  \label{eq:xa2}
  \text{erfc}\left(\frac{w_a(t)-x_0}{ \sqrt{4D_\parallel t}}\right)
  -\text{erfc}\left(\frac{w_a(t)+x_0}{\sqrt{4D_\parallel t}}\right)
  = 1\,.
\end{align}
From this expression, we can solve for the typical altruist wealth numerically
and the result is shown in Fig.~\ref{fig:xt}.

For individualists, the wealth distribution of the first individualist to go
bankrupt is given by
\begin{align}
  G_1(x,t) = S_1(t)\,G(x,t)\,,
\end{align}
We can once again obtain the evolution of the typical wealth of this first
individualist, $w_1(t)$, by applying criterion \eqref{eq:typg1} to
$G_1(x,t)$.  This gives the implicit equation for $w_1(t)$:
\begin{align}
  \text{erfc}\left(\frac{w_1(t)-x_0}{ \sqrt{4Dt}}\right)
  -\text{erfc}\left(\frac{w_1(t)+x_0}{\sqrt{4Dt}}\right)
  =  \frac{\displaystyle \text{erf}\left(\frac{x_0}{\sqrt{4\,D\,t}}\right)}
  {\displaystyle S_1(t)}
  = \frac{1}{\displaystyle \text{erf}\left(\frac{x_0}{\sqrt{4\,D\,t}}\right)}\,.
\end{align}
Similarly, the distribution of the wealth of the second individualist to go
bankrupt is
\begin{align}
  G_2(x,t) = S_2(t)\,G(x,t)\,.
\end{align}
Applying criterion \eqref{eq:typg1} to $G_2(x,t)$ yields the implicit
equation for $w_2(t)$, the typical wealth of this second individualist:
\begin{align}
  \text{erfc}\left(\frac{w_2(t)-x_0}{\sqrt{4\,D\,t}}\right)
 -\text{erfc}\left(\frac{w_2(t)+x_0}{\sqrt{4\,D\,t}}\right)
  =\frac{\displaystyle
  \text{erf}\left(\frac{x_0}{\sqrt{4\,D\,t}}\right)}{\displaystyle S_2(t)}\,.
\end{align}

The numerical solutions for the typical wealth of the individualists are also
shown in Fig.~\ref{fig:xt}.  As in the case of the survival probability, the
typical altruist wealth is intermediate to that of the two individualists.
Notice also that the wealth of the longer-lived individualist initially
increases before being inexorably drawn toward bankruptcy.  This
non-monotonicity arises because the typical time for the bankruptcy of the
longer-lived individualist is larger than the diffusion time $x_0^2/D$.  For
the wealth trajectory to not reach the origin within this time period, the
trajectory must initially move \emph{away} from the origin.  Related types of
effective repulsion phenomena have been found to arise from a variety
physically motivated constraints on first-passage
trajectories.\cite{bhat2015intermediate,meerson2019large,smith2019geometrical,meerson2019geometrical}

\subsection{$N$ agents}

\begin{figure}[ht]\centering
\subfigure[$N=4$]{\includegraphics[width=0.48\textwidth]{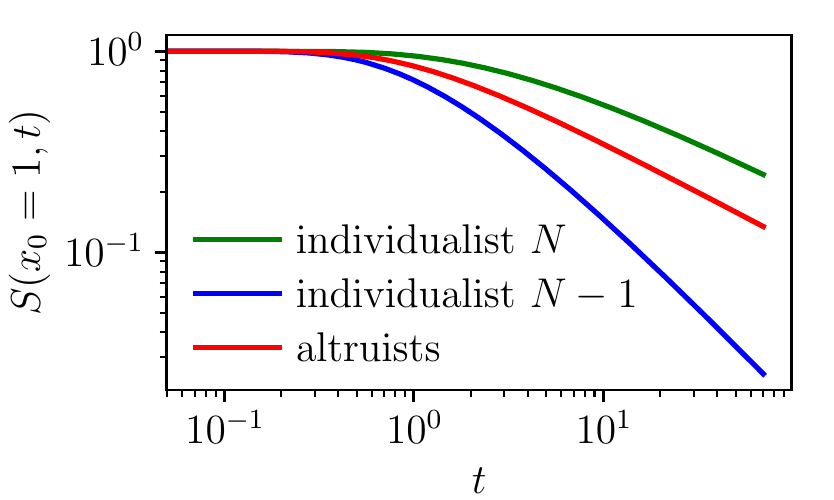}}
\subfigure[$N=16$]{\includegraphics[width=0.48\textwidth]{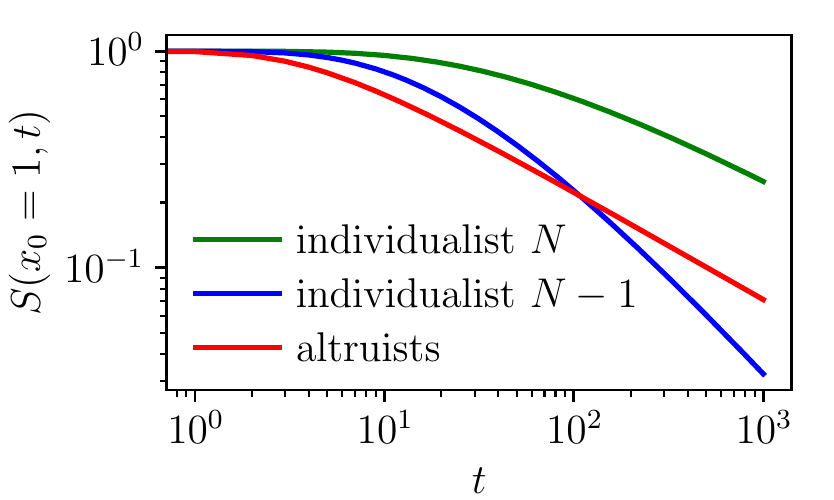}}
\caption{The survival probability of $N$ altruists (red), the
           survival probability of the last of the $N$ individualist to go bankrupt
           (green), and the one before last individualist to go bankrupt (blue).
           All agents have initial wealth $x_0=1$.}
\label{fig:SN}
\end{figure}

The calculations in the previous section can be straightforwardly extended to
$N$ agents.  The survival probability of $N$ altruists is given by
\begin{align}
  S_a(t) = S(x_0,t,D/N)\,.\label{eq:Sna}
\end{align}
with $S(x_0,t,D)$ from \eqref{eq:S}.  The diffusion coefficient of the
effective particle is now $D/N$ because the projection of a displacement in a
coordinate direction onto the diagonal $(1,1,\ldots,1)$ is $1/\sqrt{N}$ . The
survival probability of the $n^\text{th}$ individualist to go bankrupt is
given by
\begin{align}
  \label{eq:Sn}
  S_n(t) = \sum_{m=0}^{n-1} \binom{N}{m}\,\text{erfc}\left(\frac{x_0}{\sqrt{4\,D\,t}}\right)^{m}\,\text{erf}\left(\frac{x_0}{\sqrt{4\,D\,t}}\right)^{N-m}\,.
\end{align}
where the combinatorial factor accounts for the fact that there are
$\binom{N}{m}$ possible groups of $m$ bankrupt individualists among $N$, with
$m<n$, when the $n^\text{th}$ individualist is alive.  We compare the
survival probabilities of last and one before last individualists to go
bankrupt with the altruists for $N=4$ and $16$ in Fig.~\ref{fig:SN}. We see
that for large times, the survival probability of the altruists is larger
than all but the most long-lived individualist.

For $t\to\infty$, the asymptotic decay of the altruist survival probability
\eqref{eq:Sna} and the survival probabilities of each of the
individualists \eqref{eq:Sn} is
\begin{align}
\label{eq:decaySna}
\begin{split}
  S_a(t) &\sim \frac{x_0\sqrt{N}}{\sqrt{\pi D t}}\\[2mm]
  S_n(t) &\sim \binom{N}{n-1} \left(\frac{x_0}{\sqrt{\pi D t}}\right)^{N-n+1}\,.
\end{split}  
\end{align}
To obtain the asymptotics of $S_n(t)$, we use the fact that the sum in
\eqref{eq:Sn} is dominated by the term with $m=n-1$, and that the asymptotic
form of the error function is $\text{erf}(x) \sim 2x/\sqrt{\pi}$ for
$x\rightarrow 0$.  Thus the individualist survival probabilities in
\eqref{eq:decaySna} all decay asymptotically faster than $t^{-1/2}$,
\emph{except} for the $n=N$ individualist.  Here, the asymptotic form of $S_N(t)$ is
\begin{align}
  \label{eq:SnN}
  S_N(t) &\sim \frac{x_0 N}{\sqrt{\pi D t}}\,,\quad t\rightarrow\infty\,.
\end{align}
As in the case of $N=2$, the last individualist to go bankrupt has the
largest survival probability, while the survival probability of all the other
individualists decays faster than that of the altruists.  
  
  \begin{figure}[ht]\centering
\subfigure[$N=4$]{\includegraphics[width=0.48\textwidth]{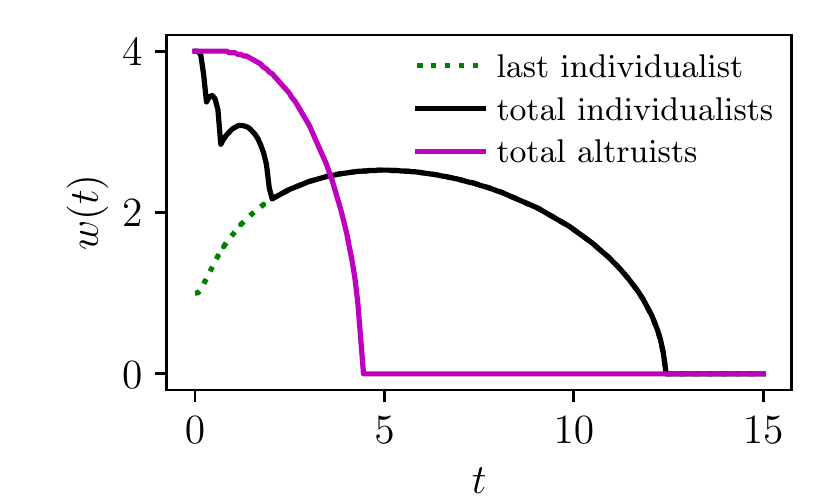}}
\subfigure[$N=16$]{\includegraphics[width=0.48\textwidth]{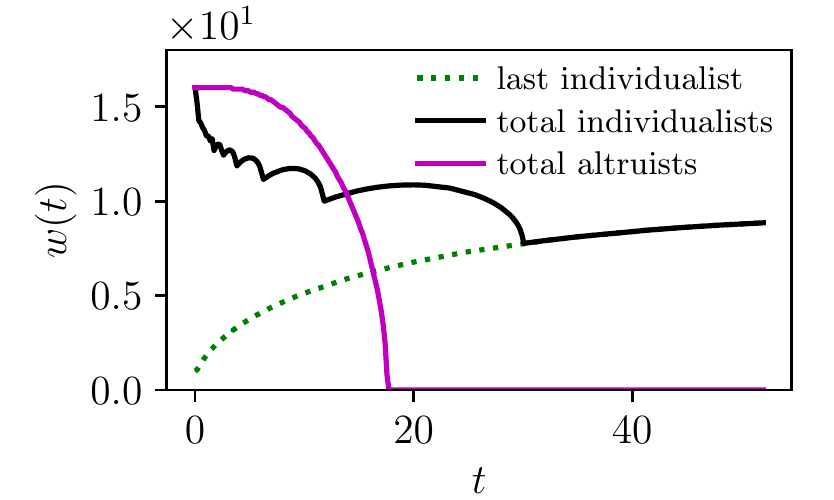}}
\caption{Evolution of the typical total wealth of $N$ altruists (violet) and
  $N$ individualists (black), and the wealth of the last individualist to go
  bankrupt for $N=4$ and $16$. All agents have initial wealth $x_0=1$.}
\label{fig:xtN}
\end{figure}
  
We may also compute the typical wealth of the agents using the same approach
as in the case of two altruists.  Now the distribution of wealth of $N$
altruists is
\begin{align}
  G_a(x,t) = S_a(t)\,G(x,t)\,,\label{eq:Gan}
\end{align}
with $G(x,t)$ given in \eqref{eq:G}. The evolution of the typical wealth
$w_a(t)$ of the $N$ altruists evolves according to Eq.~\eqref{eq:xa2}, but
with $D_\parallel$ now equal to $D/N$.  The distribution of wealth of the
$n^\text{th}$ individualist is
\begin{align}
  G_n(x,t) = S_n(t)\,G(x,t)\,,\label{eq:Gn}
\end{align}
where $G(x,t)$ again given in (\ref{eq:G}).  The evolution of the typical
wealth $w_n(t)$ of the $n^\text{th}$ individualist evolves according to
\begin{align}
  \text{erfc}\left(\frac{w_n(t)-x_0}{ \sqrt{4\,D\,t}}\right)-\text{erfc}\left(\frac{w_n(t)+x_0}{
   \sqrt{4\,D\,t}}\right) =  \frac{\displaystyle \text{erf}\left(\frac{x_0}{\sqrt{4\,D\,t}}\right)}{S_n(t)}\,.
\end{align}
The numerical comparison of the typical wealth for $N=4$ and $16$ altruists and
the first and last of the $N$ individualists, along with the total wealth in
both societies, is presented in Fig.~\ref{fig:xtN}.  We see that the first
individualist rapidly goes bankrupt, while the last individualist accumulates
the total wealth of society and survives for a long time. In contrast, the
altruistic society sees its wealth decrease monotonically with a bankruptcy
time that is intermediate to that of the first and the last individualists.

\section{Discussion}
\label{sec:disc}

We explored the role of redistribution in a toy model of wealth evolution, in
which the wealth of each person in a population of $N$ agents evolves by free
diffusion.  As a preliminary, we first studied the outcome for
individualists---these are agents that do not engage in any wealth sharing
when someone goes bankrupt.  Even though the \emph{average} wealth is
conserved during the evolution, including bankruptcy events, the
\emph{typical} wealth decreases systematically with time.  Thus
individualists go bankrupt one by one until nobody remains solvent.  For
large $N$, the first few individualists that go bankrupt suffer a harsh fate,
as their typical wealth decays rapidly with time.  Conversely, the last few
individualists to go bankrupt initially have a favorable economic outcome as
their wealth grows appreciably at early times.  This initial wealth growth
stems from an effective repulsion of their Brownian paths from the origin
(bankruptcy) because the time until their individual bankruptcies are much
larger than the diffusion time $x_0^2/D$.  To avoid bankruptcy over such a
long time period, the wealth trajectory must initially be repelled by the
origin. In fact, the most long-lived individualists typically accumulate the
total wealth of his society.  Nevertheless, every individualist ultimately
suffers the same fate of bankruptcy.

In contrast, a population of altruists equally share their wealth each time
an individual goes bankrupt.  Again, the average wealth is conserved
throughout the dynamics, but the typical wealth also systematically decreases
with time.  Thus eventually a population of altruists collectively goes
bankrupt, in which the total wealth of the population falls below a small
threshold value.  We showed that at early times altruists have a better
economic fate than individualists.  At long times, however, an
individualistic society becomes extremely inequitable, with most
individualists quickly reaching a fate of having no wealth and a few having
most of the societal wealth.  These longest-lived individualists eventually
have a better outcome than the altruists.  Thus if one is faced with a choice
of which society to join, being an average altruist is preferable to an
average individualist.

Our model is naive in many respects and there are variety of possible
extensions to consider.  The notion that an individual's wealth evolves by
free diffusion can clearly be made more realistic.  Many people draw a
regular salary, continuously spend for routine expenses, and sometimes
experience negative shocks of large unexpected expenses; this latter feature
was the focus of the study by Abebe et al.~\cite{abebe2020subsidy}.  Thus the
evolution of individual wealth is more realistically described by a process
that incorporates these features of salary, spending, and shocks.  More
realistically, the per capita wealth of societies generally increase over the
long term and it would be interesting to superimpose a slight a positive
drift on the wealth dynamics, perhaps different for each agent, that more
than compensates for the decrease in the typical wealth.

The equal-wealth sharing mechanism we studied is also unrealistically
idealized, and we focused on this simple rule because it lead to an
analytically tractable model.  It may be worthwhile to study more selfish
wealth sharing rules.  Perhaps such modifications of the wealth sharing rules
lead to better outcomes for both the survival probability and the average
wealth of an altruist population compared to an individualistic population.  It
should also be useful to explore possible connections between wealth sharing
rules and policies on optimal taxation and
redistribution~\cite{piketty2014optimal}.

\section{Acknowledgments}
SR gratefully acknowledges partial financial support from NSF grant
DMR-1910736. BD acknowledges the financial support of the Luxembourg National
Research Fund (FNR) (App. ID 14548297).

\bibliographystyle{iopart-num}


\providecommand{\newblock}{}

\end{document}